\documentclass[twocolumn]{article}

\usepackage{amssymb,amsmath}
\usepackage{subcaption}
\usepackage{graphicx}
\usepackage{authblk}
\usepackage[authoryear]{natbib}

\def\bc{\mathbf c}
\def\bd{\mathbf d}

\def\TC{\mathbf C}

\def\by{\mathbf y}
\def\TK{\mathbf K}
\def\TD{\mathbf D}
\def\by{\mathbf y}
\def\TY{\mathbf Y}
\def\TR{\mathbf R}

\def\bc{\mathbf c}

\def\TQ{\mathbf Q}
\def\TZ{\mathbf Z}

\begin{document}

\title{Does shallow geological knowledge help neural-networks to predict deep units?}

\author[1]{Bas Peters}
\author[1]{Eldad Haber}
\author[2]{Justin Granek}

\affil[1]{University of British Columbia, Vancouver, Canada}
\affil[2]{Computational Geosciences Inc.}
\date{}
\maketitle

\begin{abstract}
Geological interpretation of seismic images is a visual task that can be automated by training neural networks. While neural networks have shown to be effective at various interpretation tasks, a fundamental challenge is the lack of labeled data points in the subsurface. For example, the interpolation and extrapolation of well-based lithology using seismic images relies on a small number of known labels. Besides well-known data augmentation techniques, as well as regularization of the network output, we propose and test another approach to deal with the lack of labels. Non learning-based horizon trackers work very well in the shallow subsurface where seismic images are of higher quality and the geological units are roughly layered. We test if these segmented and shallow units can help train neural networks to predict deeper geological units that are not layered and flat. We show that knowledge of shallow geological units helps to predict deeper units when there are only a few labels for training using a dataset from the Sea of Ireland. We employ U-net based multi-resolution networks, and we show that these networks can be described using matrix-vector product notation in a similar fashion as standard geophysical inverse problems. 
\end{abstract}

\section{Introduction}

Geological interpretation of seismic images can be very time-consuming due to the large data volumes. During the last few years, new neural network designs have shown that they can provide high-quality interpretations. In particular, U-nets were used for the interpretation of geological units \citep{peters2019automatic,peters2019neural}, horizons \citep{peters2018multi}, salt \citep{zeng2018automatic}, and faults \citep{wu2019faultseg3d,Z10.1093/jge/gxy015}. 

The success of most of these methods hinges on the availability of numerous high-quality training labels. A label in seismic interpretation context is a fully or partially interpreted seismic image, or annotated pixels in the seismic image. Obtaining the labels is the most challenging part of network-based seismic interpretation. Often there are sparsely distributed labels, for example, lithology in a few boreholes, or a small number of horizon picks. Traditionally, researchers used these by training a network on small patches around the known label pixels, to predict the label at the center pixel of the patch. This approach cannot take large geological structures into account because the network never has access to full seismic images. Another approach is to manually complete the label images with sparse annotations into fully annotated images. This is very time-consuming, and the quality of the manually completed interpretation is difficult to determine.

Recently, \cite{peters2018multi,peters2019automatic} introduced partial loss-functions for non-linear regression and classification for seismic interpretation. These type of loss functions measure misfit at the known and sparsely distributed annotated label pixels only. The corresponding gradient computation also uses just the known pixels in the label images.

While partial loss functions avoid the need to work with small patches or manually generate full label images, an insufficient number of known label pixels is detrimental to the predictive power of a network. Besides relatively standard approaches that use data augmentation, \cite{peters2019neural} propose to use prior information about the expected network output to mitigate an insufficient number of annotated pixels. Their approach is to apply a quadratic smoothing penalty to the network output. For the segmentation of seismic images, the network output are probability maps for each class. This corresponds to the prior knowledge that a blocky segmentation result has smoothly transitioning probabilities for each class.

In this work, we propose another method to mitigate a lack of labels. Our method is fully compatible with data augmentation and network output regularization. We propose to combine limited lithology information from wells with seemingly uninteresting knowledge about shallow geological units. This type of information is relatively easy to obtain using robust horizon trackers applied to simple mostly layered shallow geology. We will show that the prediction accuracy significantly increases if we include such knowledge when training networks to segment deeper parts of seismic images based on well information. We present results on field data from the Sea of Ireland.

\section{Neural networks for seismic image guided interpolation of well-based lithology.}
\begin{figure}[!htb]
   \centering
   \begin{subfigure}[b]{1.0\columnwidth}
   	\centering
   	\includegraphics[width=1.0\columnwidth]{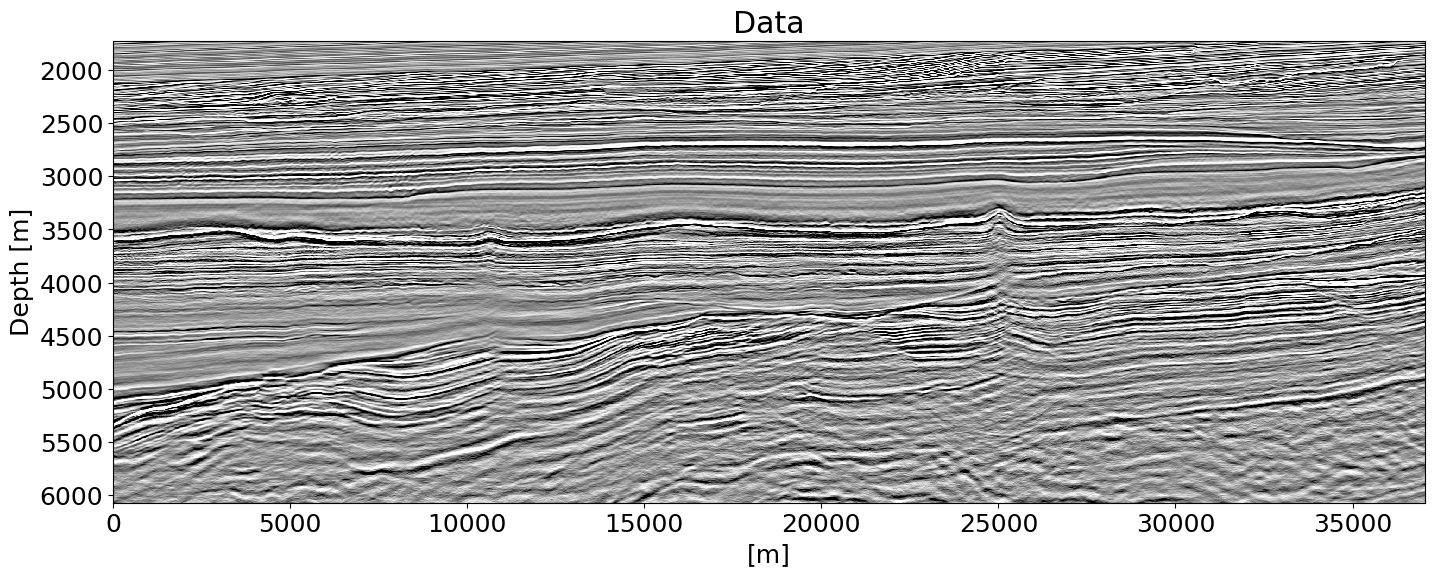}
   	\caption{}
   	\label{/SeismicLithologyWellInterp/withoutReg/data_12}
   \end{subfigure}
   \begin{subfigure}[b]{1.0\columnwidth}
   	\centering
   	\includegraphics[width=1.0\columnwidth]{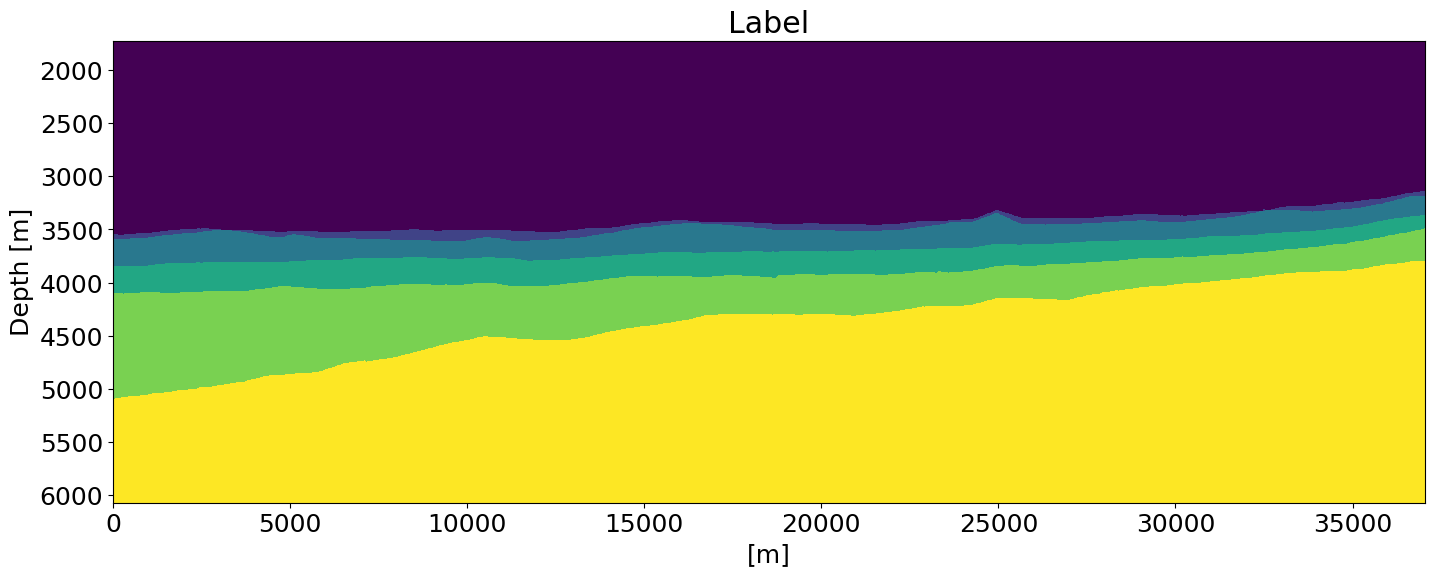}
   	\caption{}
   	\label{/SeismicLithologyWellInterp/withoutReg/label_12}
   \end{subfigure}
   \caption{(a) An example of a data image that is the input for the network. (b) A full label that we want the network to predict. Such fully annotated images are not commonly available and we never use this type of image for training a network.}
   \label{test}
 \end{figure}

\begin{figure}[!htb]
   \centering
   \includegraphics[width=\columnwidth]{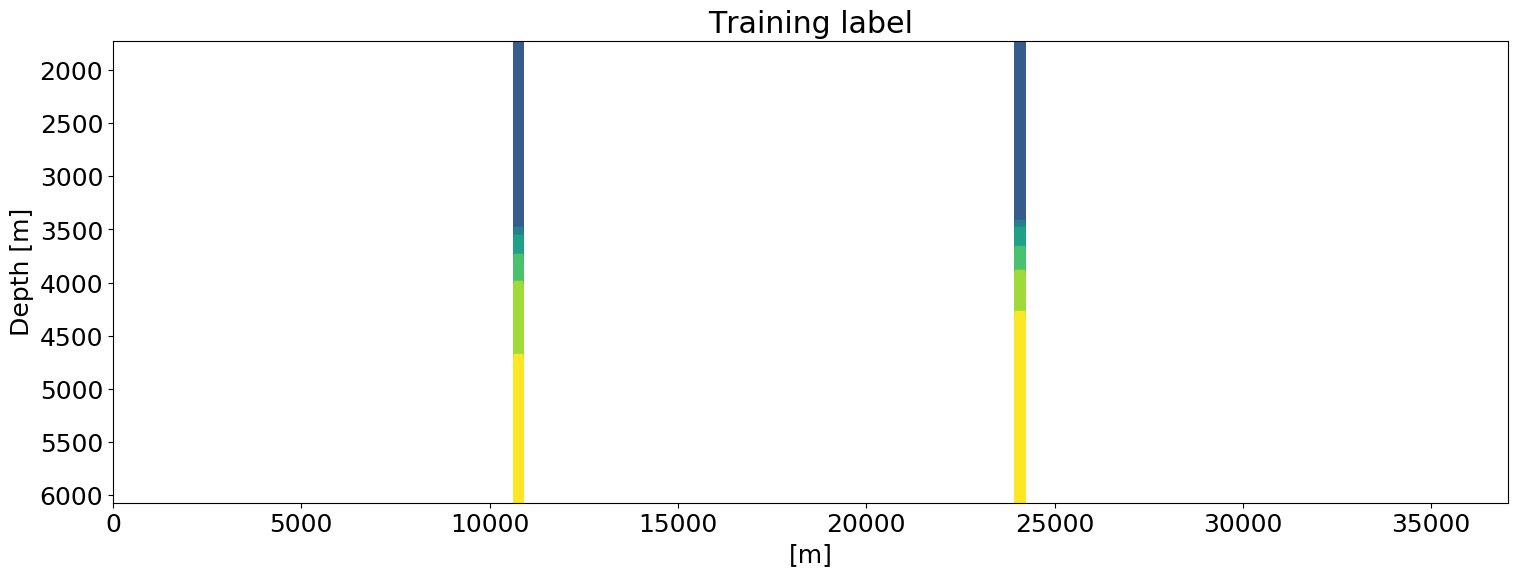}
   \caption{One of the 24 label images used for training the network. There are two columns with annotated pixels; the rest of the image is unknown and never used during training.}
   \label{/SeismicLithologyWellInterp/withoutReg/label_sub_12}
\end{figure}

The goal is to segment a seismic image as in Figure \ref{/SeismicLithologyWellInterp/withoutReg/data_12} into a number of geologically interesting units as in Figure \ref{/SeismicLithologyWellInterp/withoutReg/label_12}. Such fully annotated images are generally not available. More realistically, we have lithology information derived from well logs. This means our label images are as shown in Figure \ref{/SeismicLithologyWellInterp/withoutReg/label_sub_12}.

Our goal is thus to learn a mapping from seismic images to geologically fully segmented images while having knowledge about a small amount of well-based lithology (image columns). The network is based on the U-net \citep{Ronneberger2015}, an architecture that is successful in other semantic segmentation applications. The core of this network is an encoder-decoder structure, supplemented with skip-connections that inject high-resolution information to the up-sampled features in the decoder part of the network. This multi-resolution structure allows for the predicted of large and smaller-scale details. U-nets have also proven effective in various seismic interpretation tasks as mentioned in the introduction.

For our semantic segmentation problem, we need a neural network that maps from a seismic image of size $N = n_1 \times n_2$ to $n_\text{class}$ images that each represent the probability of a class. Inspired by \cite{treister2018low,ruthotto2018deep}, our notation and description uses matrix-vector product notation that is close to common geophysical inverse problem descriptions. We can do this by first recognizing that convolutions with small kernels are equivalent to matrix-vector products with sparse and banded matrices. Second, network states that contain multiple channels that are the result of image inputs are not very high-dimensional, and can be flattened to be written in standard linear algebraic fashion. We denote a vectorized seismic image as $\bd \in \mathbb{R}^{N}$ and any network as $f(\{\TK_i\},\bd) : \mathbb{R}^{N} \rightarrow \mathbb{R}^{N n_\text{class}}$. We will learn the network parameters $\{\TK_i\}$, which are a collection of 2D convolutional kernels of size $3\times 3$ for each layer $i$. We represent all convolutional kernels per layer in the block matrix 
\begin{equation}\label{blockconv}
\TK_i = \begin{pmatrix}
\TK_i^{1,1} & \TK_i^{1,2} &\dots & \TK_i^{1,\text{inp}} \\
\TK_i^{2,1} & \TK_i^{2,2} & \dots &  \TK_i^{2,\text{inp}}\\
\vdots         & \vdots & \ddots    & \vdots \\
\TK_i^{\text{outp},1} & \TK_i^{\text{outp},2} & \hdots & \TK_i^\text{outp,inp}
\end{pmatrix},
\end{equation}
where each sub-block is a sparse convolution matrix corresponding to a $3 \times 3$ kernel. If the block-convolution matrix is square, it means that the number of input channels is equal to the number of output channels: $\text{outp} = \text{inp}$. If the number of channels needs to increase, the corresponding block-convolution matrix is the tall matrix: $\text{outp} > \text{inp}$. A flat matrix corresponds to the case where the number of output channels is smaller than the number of input channels at a network layer $i$: $\text{outp} < \text{inp}$.

The first part of our version of a U-net follows the residual structure \citep{DBLP:journals/corr/HeZRS15} that we can also interpret as a discretization of an ordinary differential equation (ODE) \citep{HaberRuthotto2017a,Chang2017Reversible,ruthotto2018deep}. Every few layers, we subsample the image so that the network operates on multiple resolutions and different (spatial) parts of the input can influence other parts through convolutional operations, i.e., the network has the ability to learn from large-scale geological structures. Given the network input $\bd$, the initial network state is $\TY_1$, where the subscript indicates the layer number. The residual network follows as
\begin{align}
&\TY_1 = \bd \nonumber\\
&\TY_{i+1} = \TY_i - h \TK_i^\top \sigma(\TK_i  \TY_i) \quad \text{for} \quad i=2,\dots,n
\end{align}
where $h$ is equivalent to the time-step in the ODE discretization. In this notation, we represent the network state as the block vector 
\begin{equation}
\TY_i = \begin{pmatrix} \by_i^1 \\  \by_i^2 \\  \vdots \\ \by_i^{n_\text{chan}} \end{pmatrix}.
\end{equation}
If the number of channels changes, the network evolution simplifies to $\TY_{i+1} =\TR  \sigma(\TK_i  \TY_i)$ because the previous network state $\TY_i$ is of different dimensions. The matrix $\TR$ is a restriction/subsampling matrix that subsamples the network state. The nonlinear point-wise activation function, $\sigma$, is the ReLU function in our network.

At layer $n$, the network operates at the coarsest resolution. From there, we start the upsampling, or, decoder part of the network supplemented with the U-net skip connections. The network state in this second part of the network is denoted by $\TZ_i$. The network structure in the upsampling part of the network while maintaining resolution reads
\begin{equation}
\TZ_{i-1} = \TZ_i - h\TK_i \sigma(\TK_i^\top \TZ_i) \:\: \text{for} \:\: i=n-1,n-2,\dots,1
\end{equation}
If we increase the resolution, we also add the high-resolution intermediate network-state from the down-sampling arm as
\begin{equation}
\TZ_{i-1} = \sigma(\TK_i^\top  \TR^\top \TZ_i) + \TY_{i-1}.
\end{equation}
Note that we couple the convolutional kernel weights between downsampling and upsampling `arms' of the network by using the transpose kernels pairs.

The network as described above thus takes a full data image as input, and the output is a collection of full prediction images; each image is the probability map for one class. As discussed in the introduction, we want to train using partially available labels (annotated pixels in the label image) directly and without forming patches or manually completing full label images. In this way, we can learn from large-scale structure and avoid unnecessary human work. 

The block-vector $\TC \in \mathbb{R}^{n_1 n_2  n_\text{class}}_{+}$ contains one vectorized label image per class, which means the probability any given pixel corresponds to a certain class:
\begin{equation}
\TC = \begin{pmatrix} \bc^1 \\  \bc^2 \\  \vdots \\ \bc^{n_\text{class}} \end{pmatrix}.
\end{equation}
To train the network for multi-class semantic segmentation, we consider the cross-entropy loss-function, $E$, for a single data image and sum over all pixels as
\begin{equation}
L(\TD,\{\TK\},\TC) = - E(\TC , f(\{\TK\},\TD)).
\end{equation}
For our applications, most entries in $\TC$ are unknown. Therefore we need to use a loss function that only uses the known label pixels. We use the partial loss function that was introduced in a similar context by \cite{peters2019automatic}, which reads
\begin{equation}\label{partial_ce}
L(\TD,\{\TK\},\TC))_\TQ = - E(\TQ\TC, \TQ f(\{\TK\},\TD)).
\end{equation}
The matrix $\TQ$ is a subset of the the rows of the identity matrix and samples the network prediction at the known label pixel locations: $\TQ f(\{\TK\},\TD)$. The computation of this loss requires a full forward pass through the network, $f(\{\TK\},\TD)$, followed by subsampling the result. The gradient of $L(\TD,\{\TK\},\TC))_\TQ$ depends on the known label pixels only.

\begin{figure}[!htb]
   \centering
   \begin{subfigure}[b]{1.0\columnwidth}
   	\centering
   	\includegraphics[width=1.0\columnwidth]{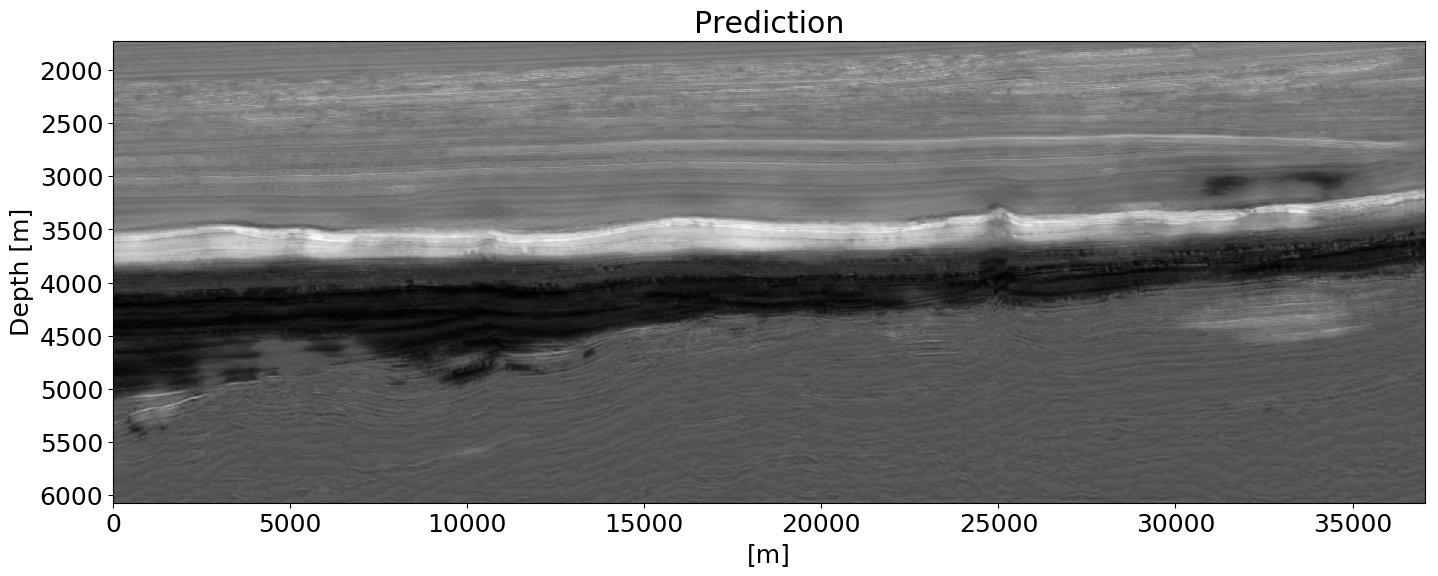}
   	\caption{}
   	\label{/SeismicLithologyWellInterp/withoutReg/prediction_12}
   \end{subfigure}
   \begin{subfigure}[b]{1.0\columnwidth}
   	\centering
   	\includegraphics[width=1.0\columnwidth]{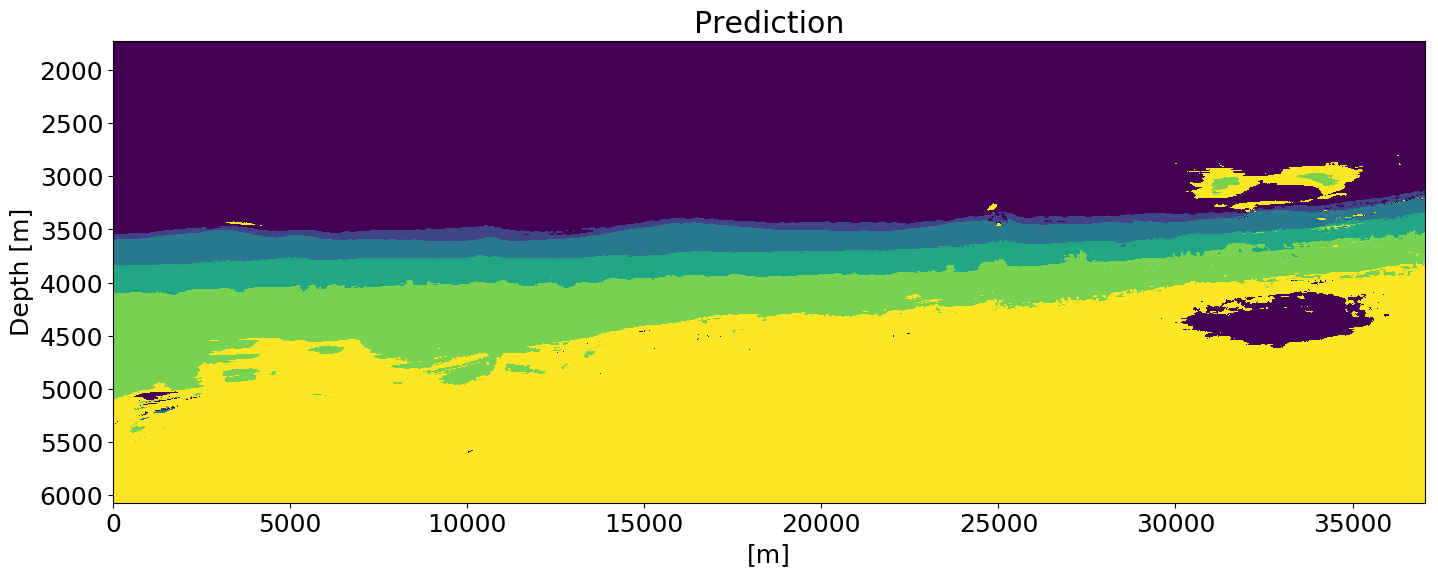}
   	\caption{}
   	\label{/SeismicLithologyWellInterp/withoutReg/prediction_threshold12}
   \end{subfigure}
   \caption{(a) Probability map for one of the classes; this is one of the outputs of the network. (b) Maximum predicted probability per pixel. Training data are $24$ seismic images with $2$ labeled columns each, see Figure \ref{/SeismicLithologyWellInterp/withoutReg/label_sub_12}.}
   \label{test1}
 \end{figure}

The training data for the network input are $24$ seismic images from a larger 3D seismic volume from the Sea of Ireland. There are also $24$ label images, where each image has only two labeled columns at well locations. We use stochastic gradient descent to train a network that has $18$ layers in the down-sampling arm, while reducing resolution six times. The upsampling arm is the same in reverse order, so there are $36$ layers in total. The number of channels varies from six up to $32$. In a similar experiment, \cite{peters2019automatic} showed that it is possible to obtain high-quality segmentations of the seismic images provided there are a sufficient number of known labeled columns. Here we will use lithology information from just two wells per image. Figure \ref{/SeismicLithologyWellInterp/withoutReg/prediction_12} shows the output of the network for one of the six classes. The final result is the maximum predicted class probability per pixel and is shown in Figure \ref{/SeismicLithologyWellInterp/withoutReg/prediction_threshold12}; not accurate almost everywhere when compared to the true label in Figure \ref{/SeismicLithologyWellInterp/withoutReg/label_12}. The true labels were provided by an industrial partner.

While neural networks are powerful tools for tasks similar to the one described in this example, the results remind us that a sufficient number of data images and labels is essential for good predictions. Acquiring additional labels would be costly as it involves drilling new wells. Data-augmentation is a relatively standard approach to get more out of the same data, starting as simple as flipping all data and label images around the vertical axis. \cite{peters2019neural} propose another method that mitigates a lack of labels by combining neural networks with well-known geophysical regularization techniques. That method applies smoothing regularization to the network output while training, such that the network class probability predictions are smooth, rather than oscillatory as in Figure \ref{/SeismicLithologyWellInterp/withoutReg/prediction_12}. 

In the following section, we present another way to mitigate a lack of known label pixels.

\section{Do known shallow geological units help to predict deeper ones?}

A small number of available labels is an inherent challenge in the Earth sciences. Techniques related to data augmentation aim to get the most out of the limited available labels. The foundation of network output regularization is injecting more information into the network training by using prior knowledge. Here we introduce another method, different from data augmentation of regularization, to train networks with greater prediction power. 

\begin{figure}[!htb]
   \centering
   \includegraphics[width=\columnwidth]{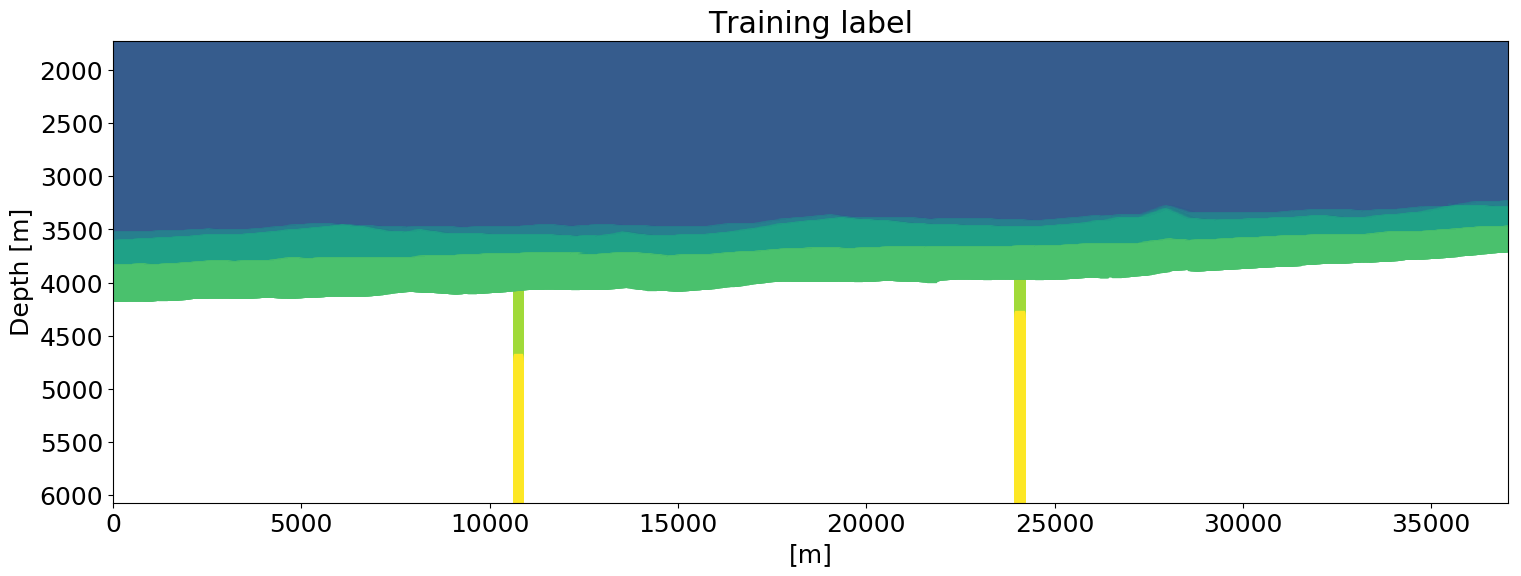}
   \caption{A label image based on information from wells and a shallow geological units that are not the target of interest. The white space is never used by the network or loss-function.}
   \label{/WellPlusLayer_NoReg/label_sub_12}
\end{figure}

The core idea is as follows: there are often one or more shallow geological units that can be delineated from just one or a few seed points, accurately and fast, using auto-tracking algorithms that are not based on learning, e.g., \citep{doi:10.1190/geo2017-0830.1}. While these shallower geological units are not the targets of interest, they do carry information about the subsurface. Figure \ref{/SeismicLithologyWellInterp/withoutReg/label_12}, for example, shows that a somewhat wedge-shaped unit separates the geological units at the top and bottom. This is a relatively simple shape (aside from the small but important deviations), so information from the two wells in addition to known top units may be helpful. The label images for this case are thus as shown in Figure \ref{/WellPlusLayer_NoReg/label_sub_12}. 

\begin{figure}[!htb]
   \centering
   \begin{subfigure}[b]{1.0\columnwidth}
   	\centering
   	\includegraphics[width=1.0\columnwidth]{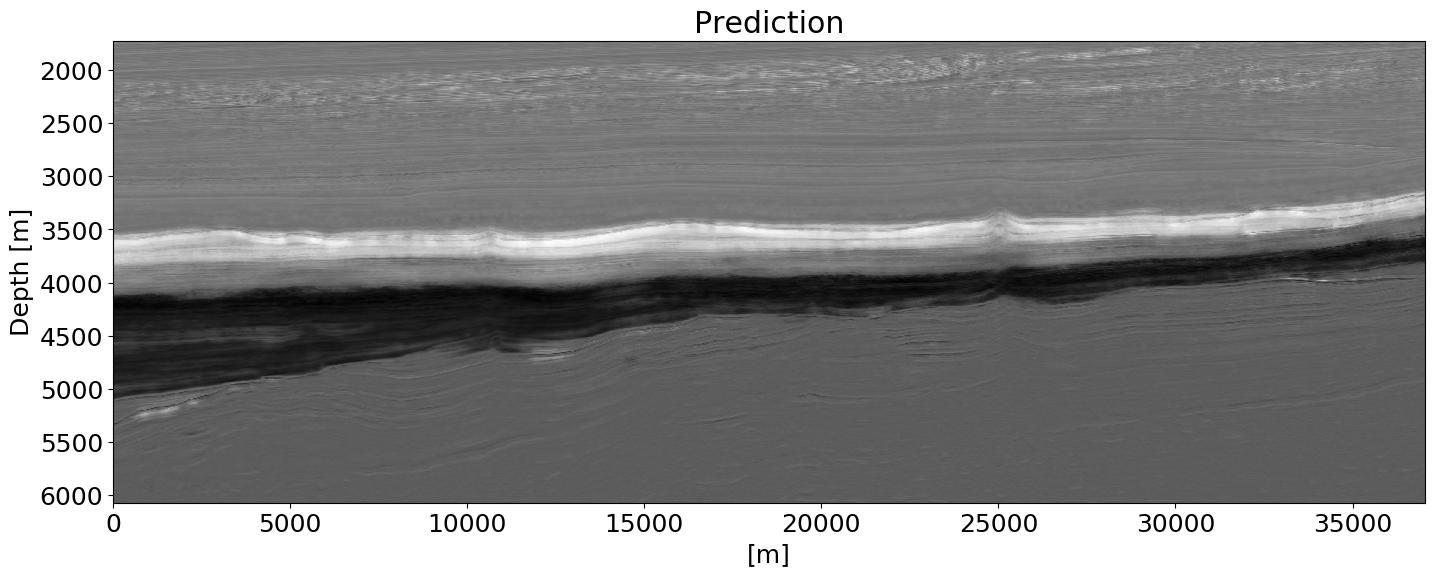}
   	\caption{}
   	\label{/WellPlusLayer_NoReg/prediction_12}
   \end{subfigure}
   \begin{subfigure}[b]{1.0\columnwidth}
   	\centering
   	\includegraphics[width=1.0\columnwidth]{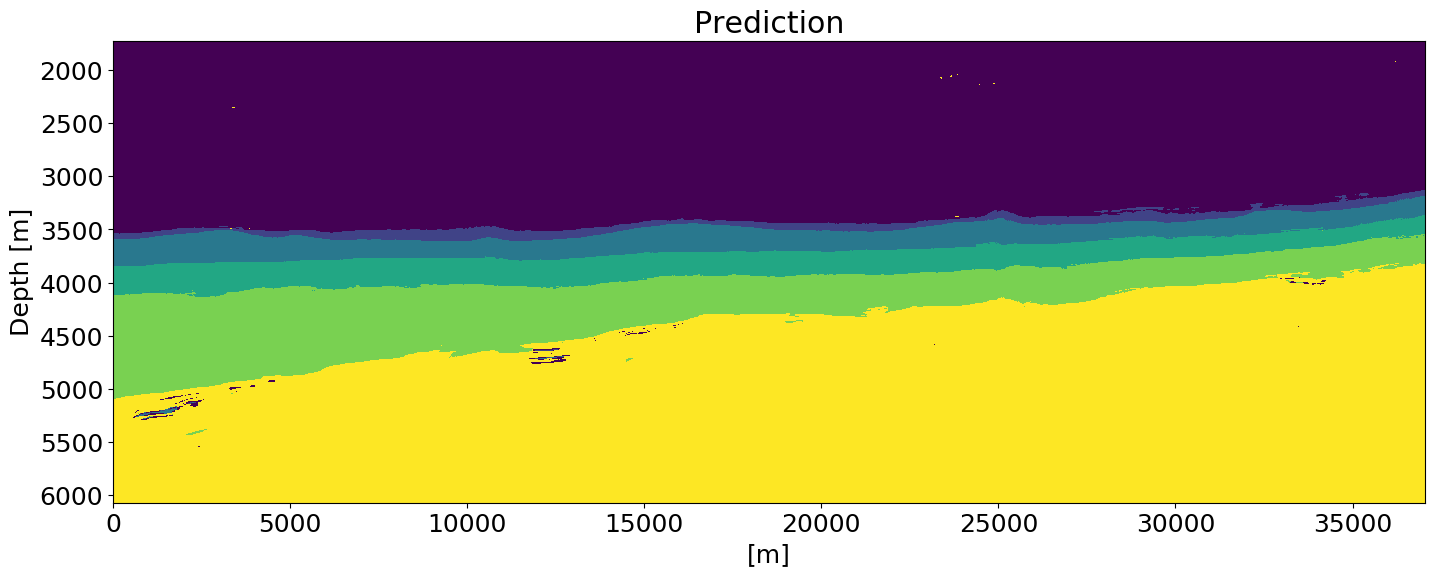}
   	\caption{}
   	\label{/WellPlusLayer_NoReg/prediction_threshold12}
   \end{subfigure}
   \caption{Prediction from a network trained on labels that use well-based information supplemented by knowledge on the shallow geology. See Figure \ref{/WellPlusLayer_NoReg/label_sub_12} for an example of a label.}
   \label{test3}
 \end{figure}

We test if knowledge of a few shallow and almost flat layers helps us predicting deeper geological units, using the same network as in the previous sections. Figures \ref{/WellPlusLayer_NoReg/prediction_12} and \ref{/WellPlusLayer_NoReg/prediction_threshold12} show the predicted probability for one of the classes and the thresholded prediction per pixel respectively. These results are much improved compared to well-based labels only in Figure \ref{/SeismicLithologyWellInterp/withoutReg/prediction_threshold12}.  

While the results show that we benefit from including knowledge about the shallow geology, researchers and practitioners in machine learning and seismic interpretation would like to know why certain methods increase the performance of a network and what task the network really learned to perform? Although providing solid theoretical guarantees is extremely difficult, we can still illuminate some network properties and concepts from a geophysical point of view to enhance our intuition.

The first observation is that including labels about the shallow geology provides us with examples of what our units of interest do not look like. This helps to prevent the network to predict our deep units of interest in shallow locations falsely. Vice versa, because the network knows how to detect the shallow units correctly, it rarely predicts the true shallow units in deep places. So even though we did not include additional information or labels about the targets of interest (the two deepest units), providing extra labels about everything else has a similar effect on the predictive capabilities of the network. Of course, the effectiveness of this approach assumes we that there are some geological units in the shallow subsurface that are relatively easy to delineate with high accuracy.

The concept described in the previous paragraph has some striking similarity with other geophysical problems like seismic full-waveform inversion. Consider a seismic imaging problem where we want to image deep targets. The waves propagate through the entire model, from shallow to deep and back to the receivers at the surface. While we want to image the deep part, we also need to obtain a good model estimate in the shallow part of the model. To image the shallow subsurface, we typically include source-receiver pairs such that the waves between those two do not travel through the deep parts of the model. The analogy with training neural networks follows from considering the forward propagation through the network as forward wave propagation. The addition of labeled pixels in the shallow part as in Figure \ref{/WellPlusLayer_NoReg/label_sub_12}, is similar to adding observed data from source-receiver pairs that record waves that propagated in the near-surface. In waveform inversion, we compute the misfit at the receiver locations only, and back-propagate the wavefield to compute update directions. Our partial loss function does exactly the same: we measure the misfit at known label pixels only, after which we compute the gradient using the (network) back-propagation algorithm to update the convolutional kernels. A difference is that the primary interest of waveform inversion are the model parameters themselves (e.g., a velocity model). In neural-network applications, the model parameters (convolutional kernels) are just a tool to obtain good predictions for the labels.

Given the similarity between geophysical inverse problems and the training of neural networks, we feel that concepts like partial loss functions, regularization, and adding label information on things other than our primary interest is not ad-hoc trial-and-error experimentation, but has a well-established base in geophysical inverse problem practice and theory.

\section{Conclusions}
We presented a new method to mitigate the detrimental effects of an insufficient number of labels when training neural networks to segment seismic images into geological units. We assume that horizon auto tracking algorithms that are not based on learning can provide quick and reliable delineated geological units in the shallow subsurface when the sedimentary layers are mostly flat. We can then use this knowledge to supplement sparse borehole-based labels in the deeper parts of the seismic images, where the image quality is lower and geological units are not flat. Results on a dataset from the Sea of Ireland verify that the shallow geological knowledge helps neural networks to segment deeper parts in the seismic image. This method may be used together with data augmentation and regularization of the network output to successfully apply neural networks to geophysical data interpretation when we do not have many labels in the subsurface. Furthermore, we described the similarities of this idea with familiar geophysical inverse problems. Finally, we show that we can also use standard sparse matrix-vector notation to describe the particular network design that we used; a variant of the U-net.

\onecolumn
\bibliographystyle{abbrvnat}
\bibliography{seismic_horizon}

\begin{thebibliography}{13}
\providecommand{\natexlab}[1]{#1}
\providecommand{\url}[1]{\texttt{#1}}
\expandafter\ifx\csname urlstyle\endcsname\relax
  \providecommand{\doi}[1]{doi: #1}\else
  \providecommand{\doi}{doi: \begingroup \urlstyle{rm}\Url}\fi

\bibitem[Chang et~al.(2018)Chang, Meng, Haber, Ruthotto, Begert, and
  Holtham]{Chang2017Reversible}
B.~Chang, L.~Meng, E.~Haber, L.~Ruthotto, D.~Begert, and E.~Holtham.
\newblock Reversible architectures for arbitrarily deep residual neural
  networks.
\newblock In \emph{AAAI Conference on AI}, 2018.

\bibitem[Haber and Ruthotto(2017)]{HaberRuthotto2017a}
E.~Haber and L.~Ruthotto.
\newblock Stable architectures for deep neural networks.
\newblock \emph{Inverse Problems}, 34\penalty0 (1):\penalty0 014004, dec 2017.
\newblock \doi{10.1088/1361-6420/aa9a90}.

\bibitem[He et~al.(2015)He, Zhang, Ren, and Sun]{DBLP:journals/corr/HeZRS15}
K.~He, X.~Zhang, S.~Ren, and J.~Sun.
\newblock Deep residual learning for image recognition.
\newblock \emph{CoRR}, abs/1512.03385, 2015.
\newblock URL \url{http://arxiv.org/abs/1512.03385}.

\bibitem[Peters et~al.(2018)Peters, Granek, and Haber]{peters2018multi}
B.~Peters, J.~Granek, and E.~Haber.
\newblock Multi-resolution neural networks for tracking seismic horizons from
  few training images.
\newblock \emph{arXiv preprint arXiv:1812.11092}, 2018.

\bibitem[Peters et~al.(2019{\natexlab{a}})Peters, Granek, and
  Haber]{peters2019automatic}
B.~Peters, J.~Granek, and E.~Haber.
\newblock Automatic classification of geologic units in seismic images using
  partially interpreted examples.
\newblock \emph{arXiv preprint arXiv:1901.03786}, 2019{\natexlab{a}}.

\bibitem[Peters et~al.(2019{\natexlab{b}})Peters, Haber, and
  Granek]{peters2019neural}
B.~Peters, E.~Haber, and J.~Granek.
\newblock Neural-networks for geophysicists and their application to seismic
  data interpretation.
\newblock \emph{arXiv preprint arXiv:1903.11215}, 2019{\natexlab{b}}.

\bibitem[Ronneberger et~al.(2015)Ronneberger, Fischer, and
  Brox]{Ronneberger2015}
O.~Ronneberger, P.~Fischer, and T.~Brox.
\newblock U-net: Convolutional networks for biomedical image segmentation.
\newblock \emph{Medical Image Computing and Computer-Assisted Intervention –
  MICCAI 2015}, page 234–241, 2015.
\newblock ISSN 1611-3349.
\newblock \doi{10.1007/978-3-319-24574-428}.
\newblock URL \url{http://dx.doi.org/10.1007/978-3-319-24574-428}.

\bibitem[Ruthotto and Haber(2018)]{ruthotto2018deep}
L.~Ruthotto and E.~Haber.
\newblock Deep neural networks motivated by partial differential equations.
\newblock \emph{arXiv preprint arXiv:1804.04272}, 2018.

\bibitem[Treister et~al.(2018)Treister, Ruthotto, Sharoni, Zafrani, and
  Haber]{treister2018low}
E.~Treister, L.~Ruthotto, M.~Sharoni, S.~Zafrani, and E.~Haber.
\newblock Low-cost parameterizations of deep convolution neural networks.
\newblock \emph{arXiv preprint arXiv:1805.07821}, 2018.

\bibitem[Wu and Fomel(2018)]{doi:10.1190/geo2017-0830.1}
X.~Wu and S.~Fomel.
\newblock Least-squares horizons with local slopes and multigrid correlations.
\newblock \emph{GEOPHYSICS}, 83\penalty0 (4):\penalty0 IM29--IM40, 2018.
\newblock \doi{10.1190/geo2017-0830.1}.
\newblock URL \url{https://doi.org/10.1190/geo2017-0830.1}.

\bibitem[Wu et~al.(2019)Wu, Liang, Shi, and Fomel]{wu2019faultseg3d}
X.~Wu, L.~Liang, Y.~Shi, and S.~Fomel.
\newblock Faultseg3d: using synthetic datasets to train an end-to-end
  convolutional neural network for 3d seismic fault segmentation.
\newblock \emph{Geophysics}, 84\penalty0 (3):\penalty0 1--36, 2019.

\bibitem[Zeng et~al.(2018)Zeng, Jiang, and Chen]{zeng2018automatic}
Y.~Zeng, K.~Jiang, and J.~Chen.
\newblock Automatic seismic salt interpretation with deep convolutional neural
  networks.
\newblock \emph{arXiv preprint arXiv:1812.01101}, 2018.

\bibitem[Zhang et~al.(2019)Zhang, Yang, Sun, and Li]{Z10.1093/jge/gxy015}
H.~Zhang, C.~Yang, H.~Sun, and S.~Li.
\newblock {Seismic fault detection using an encoder–decoder convolutional
  neural network with a small training set}.
\newblock \emph{Journal of Geophysics and Engineering}, 03 2019.
\newblock \doi{10.1093/jge/gxy015}.
\newblock URL \url{https://doi.org/10.1093/jge/gxy015}.

\end{thebibliography}

\end{document}